\documentclass[letterpaper, 10 pt, conference]{ieeeconf}

\IEEEoverridecommandlockouts
\usepackage{cite}
\usepackage{amsmath,amssymb,amsfonts}
\usepackage{algorithmic}
\usepackage{graphicx}
\usepackage{textcomp}
\usepackage{xcolor}

\let\proof\relax

\usepackage{amsthm}
\newtheorem*{remark}{Remark}
\newtheorem{theorem}{Theorem}
\usepackage{subfigure}

\def\BibTeX{{\rm B\kern-.05em{\sc i\kern-.025em b}\kern-.08em
    T\kern-.1667em\lower.7ex\hbox{E}\kern-.125emX}}
\begin{document}

\title{\LARGE \bf Real-Time  Cross-Fleet Pareto-Improving Truck Platoon Coordination*\\

%

\thanks{*This work is supported by the Strategic Vehicle Research and Innovation Programme through the project Sweden for Platooning, Horizon 2020 through the project Ensemble, the Knut and Alice Wallenberg Foundation, the Swedish Foundation for Strategic Research and the Swedish Research Council.} 
	\thanks{$^{1}$A. Johansson and J. M\aa rtensson are with the Integrated Transport Research Lab and Division of Decision and Control Systems,
		School of Electrical Engineering and Computer Science, KTH Royal Institute
		of Technology, Stockholm, Sweden.,
		SE-100 44 Stockholm, Sweden. Email:
		{\tt\small \{alexjoha,  jonas1\}@kth.se}}
	\thanks{$^{2}$X. Sun and Y. Yin are with the Department of Civil and Environmental Engineering, University of Michigan, 2350 Hayward St, Ann Arbor, MI 48109.
	 Email: {\tt \small \{xtsun, yafeng\}@umich.edu}  }
}

\author{Alexander Johansson$^{1}$, Xiaotong Sun$^{2}$, Jonas M\aa rtensson$^{1}$ and Yafeng Yin$^{2}$}

\maketitle
\thispagestyle{empty}
\pagestyle{empty}

\begin{abstract}
This paper studies a multi-fleet platoon coordination system in transport networks that deploy hubs to form trucks into platoons. The trucks belong to different fleets that are interested in increasing their profits by platooning across fleets. The profit of each fleet incorporates platooning rewards and costs for waiting at hubs. Each truck has a fixed route and a waiting time budget to spend at the hubs along its route. To ensure that all fleets are willing to participate in the system, we develop a cross-fleet Pareto-improving coordination strategy that guarantees higher fleet profits than a coordination strategy without cross-fleet platoons. By leveraging multiple hubs for platoon formation, the coordination strategy can be implemented in a real-time and distributed fashion while largely reducing the amount of travel information to be shared for system-wide coordination. We evaluate the proposed strategy in a simulation study over the Swedish transportation network. The cross-fleet platooning strategy significantly improves fleets' profits compared with single-fleet platooning, especially the profits from smaller fleets. The cross-fleet platooning strategy also shows strong competitiveness in terms of the system-wide profit compared to the case when a system planner optimizes all fleets' total profit. 
\end{abstract}

\section{Introduction}

A truck platoon refers to a group of trucks driving on roads with small inter-vehicular distances, enabled by connected automated vehicle technology. Platooning has the potential to significantly reduce the environmental impact as well as trucks' operational cost. The report in \cite{Chottani2018} anticipated that platooning can produce total cost savings of trucks in the US by up to $19\%$ from 2025 to 2027. The savings are due to reduced fuel consumption and reduced workload of drivers. Reduced fuel consumption in platoon driving owes to reduced air drag. The field experiments in \cite{Browand2004, Lammert2014, Alam2015,Tsugawa2016} reported that fuel savings of following trucks to reach approximately $10\%$. Drivers in the following vehicles reduce their workloads by leaving the longitudinal control to the leading vehicle's driver \cite{shladover2015cooperative}. Truck platooning can also increase driving safety and road capacity due to the synchronized driving in platoons, which are discussed in \cite{Ioannou1993,Fernandes2012}.

\begin{figure}
    \centering
    \includegraphics[scale=0.6]{ 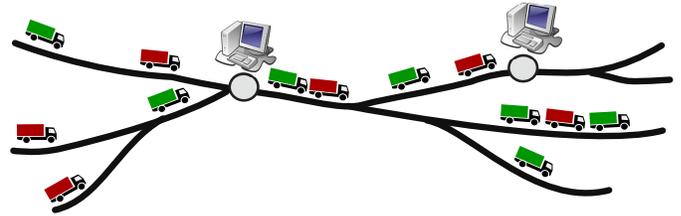}
    \caption{Network with roads, junctions, and hubs where trucks from different fleets can form platoons.}
    \label{networkex}
\end{figure}

Platooning trucks need low-level coordination to drive fuel-efficiently and safely. The interested readers are referred to \cite{Horowitz2000,Besselink2016} for an overview regarding vehicular control design and system architecture for low-level coordination. Considering the dissimilarities of truck schedules and routes in operations, high-level coordination among trucks is also needed to form platoons. This paper studies a real-time coordination problem for trucks associated with different fleets to form profitable platoons at hubs in general transportation networks.  As illustrated in Figure~\ref{networkex}, trucks can stop and wait at hubs to form platoons with others. For the approaching trucks to each hub, a coordinator is deployed to compute coordination solutions that specify departure times and platoon partners. Therefore, each truck's overall platooning plan is provided by a set of hubs along its path, and each hub provides event-triggered real-time coordination. Compared to a system where trucks only coordinate within the same fleet, cross-fleet coordination can largely improve platooning opportunities, leading to greater system benefits. Compared to a system where a coordinator forms platoons for all trucks to maximize total system profit, cross-fleet coordination emphasizes that each fleet participates in the system to increase its profit under the possibility of cooperating with other fleets.

\subsection{Related work on platoon coordination}

Many of the previous works on platoon coordination strategies have been undertaken from a centralized system perspective. These studies aim to maximize all trucks' total profit by optimizing waiting times at a hub \cite{Zhang2017,Boysen2018, Larsen2019,Pourmohammad2020} or adjusting speeds en-route \cite{ Larsson2015,Liang2016, Hoef2018, abdolmaleki2019itinerary,Xiong2019}.  Most of them adopt static planning models of platooning, where a planner gathers trucks' travel information and makes coordination decisions before the execution of trips. Real-time coordination schemes considered in \cite{Liang2016, Hoef2018,Xiong2019} allow decisions to be made when trucks are en-route. A more detailed review of studies on systematic platoon planning and coordination strategies is given in \cite{Bhoopalam2018}. 

Platoon coordination when each truck optimizes its own profit was considered in \cite{Farokhi2013} and in our previous research efforts \cite{Johansson2018,Sun2019,Sun2021, Johansson2021_2}. The motivation lies in the fact that trucks can be owned by small transport companies or even single persons, making centralized coordination hard to implement. The works in \cite{Farokhi2013, Johansson2018,Sun2021, Johansson2021_2} use non-cooperative games to model and study strategic interactions among trucks when they form platoons to maximize their individual profits. The work in \cite{Sun2019} uses a cooperative game to design a profit redistribution mechanism to incentivize trucks on the same road segment to form and maintain the optimal platoon formation. Following this line of research, the recent work in \cite{Zeng2020} first proposes a non-cooperative game to model fleet owners' strategies when they have more than one truck to platoon with others.

\subsection{Contributions}

This paper considers a platooning system with multiple fleets, where each fleet aims to maximize its own profit through platoon cooperation. Such a platooning system is of practical importance, considering that freight carriers usually employ multiple trucks and have commercial interests in optimizing their own revenues. Though a similar problem setting has been considered in \cite{Zeng2020}, many related open research questions are needed to be extensively studied. For instance, the non-convexity of the proposed platoon coordination game in \cite{Zeng2020} imposes impossibility to derive the optimality conditions for best responses. The given iterative algorithm then cannot guarantee the derived outcome is a Nash equilibrium. Instead, we propose a Pareto-improving coordination strategy to ensure that no fleet will be worse-off by joining the system than conducting single-fleet coordination, while avoiding the complexity in deriving Nash equilibrium solutions. Practically, such a strategy can be easily implemented under the hub-based platooning system. 

In more details, we assume that trucks have different origins, destinations, departure times, and fleet affiliations in a general transportation network. Trucks form platoons at hubs, which, for example, can be parking or rest areas along highways. Such a setting is beneficial from several perspectives. First, trucks can share information with local information gathering agents at hubs instead of the direct sharing of routing information among fleets, which is usually sensitive in the operations of commercial vehicles \cite{Cambridge2013}. As multiple hubs exist, trucks only need to share with each hub the travel information on the path segment until the next visiting one to further limit the degree of information sharing. Meanwhile, using multiple hubs lays the foundation of real-time coordination as it decomposes the complexity of the platoon coordination problem in the general network. Lastly, compared to the scenario that trucks form platoons on roads by speeding up or slowing down to merge, forming platoons at hubs vastly decreases disturbances to regular traffic. 

The main contributions of this paper are summarized as follows: 
\begin{itemize}
    \item We study a multi-fleet platoon coordination system where each of the fleets aims to maximize their own profit.
    \item We develop an event-triggered dynamic coordination process that can be implemented in real-time. 
     \item We evaluate the proposed cross-fleet Pareto-improving coordination strategy in terms of solution efficiency by comparing it with single-fleet and system-maximum coordination strategies.
     \item We conduct a simulation study to evaluate the viability of the coordination process and show that cross-fleet platooning can significantly increase fleets' revenues and reduce trucks' environmental impact.
\end{itemize}

The outline of this paper is as follows. The basic setting of the platoon coordination system is introduced in Section \ref{SMpi}.  In Section \ref{CSpi}, we describe the derivation of feasible platoons of a set of approaching trucks to a hub, and the platooning profit per fleet, which is used as input for coordination strategies established in Section \ref{CS2pi}.  A simulation study is presented in Section \ref{SSpi}, where we evaluate the gain of cross-fleet platooning. Conclusions and future work are given in Section \ref{CFpi}.

\section{System model}\label{SMpi}
This section first mathematically describes the basic model setting, including the studied transportation network and trucks' travel information. Then we introduce how the overall coordination is conducted dynamically through an information-sharing process and a sequence of coordination instances triggered under certain conditions. 
\subsection{The basic settings}

A directed graph, denoted as $\mathcal G=(\mathcal V, \mathcal E)$, is used to represent the transportation network. The set of nodes $\mathcal V$ includes road junctions and hubs, and the set of edges $\mathcal E$ includes roads that connect the nodes. 

The set of trucks is denoted as~$\mathcal N=\{1,...,N\}$, where $N$ is the number of trucks in the network. The set of fleets is denoted by $\mathcal F=\{1,...,F\}$ with $F$ being the number of fleets. The set of trucks in fleet $f$ is denoted $\mathcal N^f \subseteq \mathcal N$, with $\bigcup_{f\in \mathcal F} \mathcal N^f=\mathcal N,\: \mathcal N^f\cap \mathcal N^{f'}=\emptyset,\:\forall f\neq f'\in \mathcal F$. Each truck $i\in \mathcal N$ has a predefined path $\mathcal P_i \subseteq \mathcal E$ in the transportation network connecting its origin and destination. Path $\mathcal P_i$ can be further decomposed into a set of path segments separated by hubs: $\mathcal P_i=~\{\mathcal P_i^1,\dots,\mathcal P_i^h,\mathcal P_i^{h+1},\dots,\mathcal P_i^{H_i}\}$, where $H_i$ is the number of segments in truck $i$'s path. Each path segment starts from a hub or from truck $i$'s origin and ends at a hub or at the truck's destination. For instance, $\mathcal P_i^h$ starts at hub $h$ and ends at hub  $h+1$. The intermediate nodes in each path segment are junctions. With a slight abuse of notation, hub $h$ will later refer to a hub's global hub index instead of the $h$th hub in a truck's path.  Each truck has a waiting time budget that it can spend at hubs, denoted as $\bar w_i,\:\forall i\in \mathcal N$. Trucks depart from their origins per their schedules and arrive at their destinations within the time horizon~$T$. All trucks travel on edges with a constant speed~$v$.

\subsection{Information-sharing}
\begin{figure}
    \centering
    \includegraphics[scale=0.8]{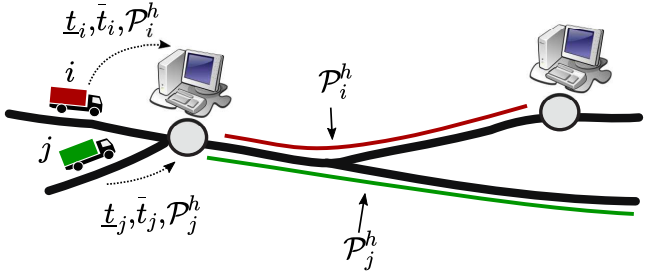}
    \caption{Trucks share information with their next hub. Truck $i$ share its arrival time, $\underline t_i$, its latest departure time, $\overline{t}_i$, and the subsequent path segment,~ $\mathcal P_i^h$. }
    \label{Shareexample}
\end{figure}

Each truck is assumed to share travel information with the next hub along its path once it departs from the current hub or its origin. In this way, trucks only share information with one hub at a time. As illustrated in Figure~\ref{Shareexample}, truck~$i$ informs hub $h$ about its arrival time at the hub, $\underline t_i$, its latest possible departure time from the hub, $\overline{t}_i$, and the path segment after hub $h$, $\mathcal P_i^h \subseteq \mathcal P_i$. Specifically, truck $i$'s latest possible departure time is limited by how much of its waiting budget has been spent at previous hubs along its path. 

\subsection{Triggering of platoon coordination}

\begin{figure}
    \centering
    \includegraphics[scale=0.8]{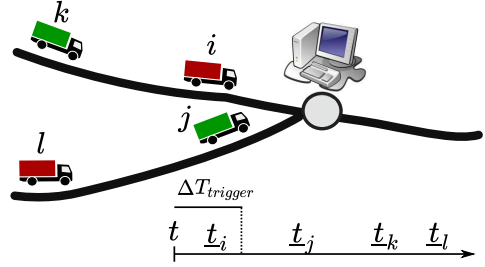}
  \caption{Illustration of the coordination triggering process. Trucks $i,j,k,l$ are approaching the hub and coordination is triggered at time $t$ since truck $i$ is uncoordinated and will arrive within $\Delta T_{trigger}$ time units. }
    \label{triggerexample}
\end{figure}

Since travel information arrives at a hub dynamically in the system, the coordination process at a hub is decomposed into a sequence of event-triggered coordination instances.  The triggering condition of one instance is when an uncoordinated truck will arrive at the hub within the next $\Delta  T_{trigger}$ time units. A truck is uncoordinated if it is not informed about its departure time at the hub. That is to say, a coordination instance at hub $h$ is triggered at time $t$, if there exists some uncoordinated approaching truck $i$ whose arrival time satisfies $t\geq \underline t_i-\Delta T_{trigger}$.  The coordination triggering  is illustrated in Figure \ref{triggerexample}.

The batch of uncoordinated trucks being considered for platooning at time $t$ is denoted as $\mathcal B_h(t)$. We select the batch size to ensure the computational time for the coordination instance is within $\Delta  T_{trigger}$ time units. The details are explained in Sections \ref{subsec_ccc_cc} and \ref{subsec_simu_ce}. Once the coordination results for trucks in $\mathcal B_h(t)$ are computed per proposed strategies, the coordinator informs the trucks who will depart in platoons about their departure times and platoon partners. They will neither trigger nor be part of future coordination at the hub. Trucks who will depart alone, according to the computed coordination results, will not be informed about their departure times and will be considered in the next batch unless they were the trucks that triggered the coordination, if its waiting budget is not zero. Trucks approaching the hub in the form of a platoon are treated as individuals in the coordination. Hence, platoons are dissembled but may be assembled again depending on the coordination results.

\section{Feasible platoons and platooning profit}\label{CSpi}

 This section first introduces the notion of the pairwise feasibility graph used to generate feasible potential platoons efficiently. Then the profit function associated with each fleet and platoon is introduced. It is composed of a platooning reward and a cost for waiting at a hub. The feasible platoons and their platooning profits are computed at each coordination instance. These are components used to formulate the optimization problems associated with the coordination strategies.

\subsection{Pairwise feasibility graph and batch selection}\label{PWFG}
 
\begin{figure}
    \centering
    \includegraphics[scale=0.75]{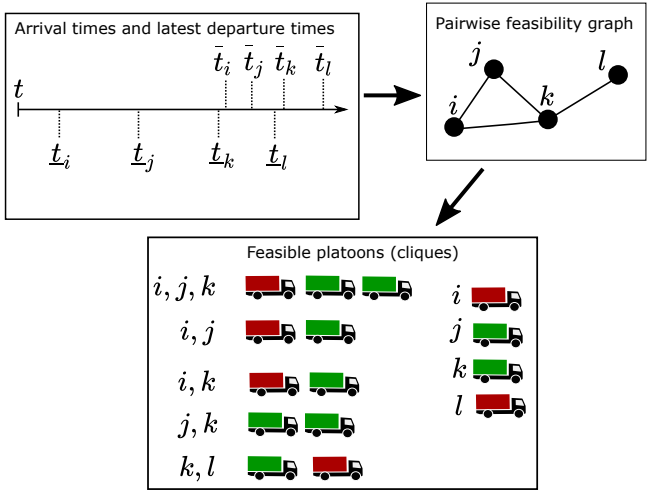}
    \caption{Illustration of the process of determining the feasible platoons from the arrival times and the latest feasible departure times. }
    \label{pwfg}
\end{figure}

To compute the feasible platoons out of a batch, we first introduce the concept of The pairwise feasibility graph. Figure \ref{pwfg} illustrates the construction of a pairwise feasibility graph and the associated feasible platoons. The nodes in the pairwise feasibility graph represent trucks. An edge between any truck pair, say truck $i$ and $j$, exists, if they have at least one common edge in their subsequent path segments and their departure time windows overlap, that is, if $\max(\underline t_i,\underline t_j)\leq \min( \overline t_i,\overline t_j)$. For a platoon to be feasible, all platoon members must be pairwise feasible to each other. Therefore, the set of feasible platoons is obtained by checking the feasibility of all cliques\footnote{In graph theory, a clique is a subset of vertices of an undirected graph such that every two distinct vertices in the clique are adjacent.} in the pairwise feasibility graph. The platoon $p$ is feasible if $\max(\underline t_i|i\in \mathcal N_p)\leq \min( \overline t_i|i\in \mathcal N_p)$, where  $\mathcal N_p \subseteq \mathcal B_h(t)$ denotes the set of trucks in the platoon $p$. If the condition above is satisfied, there must exist a departure time at hub $h$ that is feasible for all trucks in $\mathcal N_p$. If trucks in a platoon formed at a hub have different subsequent paths, then the platoon splits into sub-platoons at junctions when their paths diverge. This is captured in the platooning reward, introduced in Section \ref{PRsec}.  The approach of computing the feasible platoons of a set of trucks by the pairwise feasibility graph is similar to the approach of matching groups of riders to a fleet of shared vehicles in~\cite{Mora2017}.

 The number of trucks in the considered batch and the number of feasible platoons are crucial for the computational efficiency of the coordination problems. The batch $\mathcal B_h(t)$ is selected such that it includes less than $\bar b$ trucks and the number of feasible platoons is less than $\bar c$. This is executed by sorting the uncoordinated trucks according to their arrival time and adding one truck at a time to the batch until the number of trucks exceeds $\bar b$ or the number of feasible platoons exceeds $\bar c$. The set of feasible platoons in the selected batch  $\mathcal B_h(t)$ is denoted $\mathcal C_h(t)$. The thresholds $\bar b$ and $\bar c$ are chosen to achieve acceptable computational time for real-time coordination, which is explained in details in Sections \ref{subsec_ccc_cc} and \ref{subsec_simu_ce}.

\subsection{Platooning profit }\label{PRsec}

The profit of each platoon includes a platooning reward and a cost for waiting. The set of trucks involved in platoon $p\in \mathcal C_h(t)$ is  $\mathcal N_p = \bigcup_{f\in \mathcal F} \mathcal N_p^f$, where $\mathcal N_p^f \subseteq \mathcal B_h(t)$ is the set of trucks from fleet~$f$. As mentioned before, a platoon $p$ departing from a hub can further split into several sub-platoons when platoon members have different subsequent path segments. Therefore, we track the platooning reward per edge and fleet. The set of trucks from fleet $f$ in platoon $p$ with edge $e$ in their path segments is denoted $\mathcal N_{p,e}^f \subseteq \mathcal N_p^f$. The edges in the path segments of the trucks in $\mathcal N_{p}^f$ is denoted $\mathcal P_p^f=\bigcup_{i\in \mathcal N_{p}^f } \mathcal P_i^h$, where each edge starts and ends with a hub or a junction. Fleet $f$'s  profit associated with platoon $p$ is 
	\begin{equation*}
 r_p^f=\sum_{e \in  \mathcal P_p^f}R_e^f(\mathcal N_{p,e}^1,\dots,\mathcal N_{p,e}^F)-\Lambda^f(\mathcal N_{p}^1,\dots,\mathcal N_{p}^F),
	\end{equation*}
	where the first term is the platooning reward over the edges in the path segments and the second term is the waiting cost. The total profit of platoon $p$  is 
	\begin{equation*}	
	r_p =\sum \limits_{f\in \mathcal F} r_p^f.
		\end{equation*}
		
The simulation study presented in Section \ref{SSpi} adopts the following platooning reward function:
		\begin{equation}\label{PR}
	R_e^f(\mathcal N_{p,e}^1,...,\mathcal N_{p,e}^F)= \rho_e \ (\sum  \limits_{f\in \mathcal F} |\mathcal N_{p,e}^f|-1) \frac{|\mathcal N_{p,e}^f|}{\sum  \limits_{f\in \mathcal F} |\mathcal N_{p,e}^f|} ,
	\end{equation}
	where $\rho_e$ is the platooning profit per following truck on edge $e$, which can be customized to include length and road grade information of edge $e$ into consideration. This platooning reward function implies that  each following truck has the same platooning profit, and the platooning profit on each edge is shared among the participating fleets in proportion to the number of trucks from each fleet.

	The waiting cost function in the simulation study is specified as follows:
        \begin{equation}\label{WC}
        \Lambda^f(\mathcal N_{p}^1,...,\mathcal N_{p}^F)=	\sum \limits_{i\in \mathcal N_{p}^f} \lambda (t_p^*-\underline t_i),
    	\end{equation}
    where
    	 \begin{equation*}
        t_p^*=\max(\underline t_i|i\in \mathcal N_p)
    	\end{equation*}
    is the earliest feasible departure time of platoon $p$, and $\lambda$ is the waiting cost per time unit. This waiting cost model is accurate if the waiting cost is minimized when the platoon departs at the earliest feasible departure time and the  waiting cost of each truck is linearly increasing.

\section{Coordination strategies}\label{CS2pi}

In this section, we first present our Pareto-improving cross-fleet coordination strategy used to compute the departure times of trucks and platoons when coordination is triggered at a hub. The strategy mathematically confines feasible solutions to those that increase all fleets' individual profits by cooperation across fleets. As a comparison, we provide two other strategies that require different degrees of cooperation among fleets. The second strategy models a scenario where fleets maximize their profits without cooperating across fleets, which can be viewed as a benchmark solution. The third strategy models a scenario where fleets have a common objective of maximizing all trucks' total profit without requiring that cooperation across fleets increases all individual fleets' profits.  Last, we discuss the number of decision variables and constraints in the proposed coordination strategies' optimization problems.

\subsection{Pareto-improving cross-fleet strategy}

In this strategy, the coordinator optimizes all fleets' total profit while ensuring that each fleet obtains more profit than that without cross-fleet platoon cooperation.  The solution is a Pareto optimal solution, which is acceptable to all fleets. The Pareto-improving cross-fleet strategy is formulated as the following optimization problem: 
 \begin{subequations} \label{par}
	\begin{align}
	&\underset{\epsilon_p: p\in \mathcal C_h(t) }{\text{max}}&&  \sum \limits_{p\in \mathcal C_h(t)}  r_p \epsilon_p  & \\
	&\text{s.t.} &&
	\sum \limits_{p\in \mathcal \mathcal C_{h,i}(t) }\epsilon_p =1, & \forall i\in \mathcal B_h(t)  \label{par2} \\
	&&& \sum \limits_{p\in \mathcal C_{h}(t)}  r_p^f  \epsilon_p \geq \!  \sum \limits_{p\in  C_{h}(t)}  r_p^f  \epsilon_p',   &  \forall f\in \mathcal F_h(t) \label{par1}\\ 
	&&& \epsilon_p=\{0,1\}, & \forall p\in \mathcal C_h(t).\label{par3}
	\end{align} 
	 \end{subequations}
In this optimization problem, there is one decision variable $\epsilon_p$ for each feasible platoon $p \in \mathcal C_h(t)$. The decision variable $\epsilon_p$ equals one if platoon $p$ is used and equals zero if platoon $p$ is not used \eqref{par3}. The set $\mathcal C_{h,i}(t) \subseteq \mathcal C_h(t)$ denotes the platoons that truck $i$ is a member of, and constraints in \eqref{par2} ensure that each truck is assigned to exactly one platoon. Constraints in \eqref{par1} ensure that the solution is a Pareto improvement from the single-fleet strategy that platoon coordination happens only within each fleet. The parameter $\epsilon_p'$ is the solution obtained by the single-fleet strategy introduced in the following subsection. The set $\mathcal F_h(t)\subseteq \mathcal F $ includes the fleets involved in the batch $\mathcal B_h(t)$.  Problem \eqref{par} is ensured to admit at least one feasible solution since the single-fleet solution is feasible.


\subsection{Single-fleet strategy} \label{sfs}
In this strategy, the coordinator optimizes the fleets' profits when fleets do not cooperate in forming platoons across fleets. The single-fleet strategy for fleet $f$ is formulated as the following optimization problem:
\begin{subequations}\label{iso}
	\begin{align}
	&\underset{\epsilon_p:  p\in \mathcal C_h^{f}(t) }{\text{max}}&&  \sum \limits_{p\in \mathcal C_h^{f}(t) }  r_p^f \epsilon_p &  \\ \label{iso1}
	&\text{s.t.} &&
	\sum \limits_{p\in \mathcal C_{h,i}^{f}(t) }\epsilon_p =1, & \forall i\in \mathcal B_h^{f}(t)   \\
&&& \epsilon_p=\{0,1\},  &\forall p\in \mathcal C_h^{f}(t), \label{iso2} 
	\end{align}
	\end{subequations}
where $\mathcal C_h^{f}(t) \subseteq \mathcal C_h(t)$ denotes the platoons exclusively consisting of trucks from fleet $f$, and the set of platoons that truck $i$ is a member of is denoted $\mathcal C_{h,i}^{f}(t)$. The set $\mathcal B_h^{f}(t)$ includes the trucks from fleet $f$ in the batch $\mathcal B_{h}(t)$. Similar to the optimization problem in \eqref{par}, constraints in \eqref{iso1} ensure that each truck is assigned to exactly one platoon, and constraints in \eqref{iso2} ensure that the decision variables are binary. 
	

\subsection{System-maximum strategy} 
In this strategy, the coordinator optimizes all fleets' total profit without ensuring each of the fleets will be Pareto-improving by platooning across fleets. It is therefore  uncertain if fleets are willing to accept such a solution, different from the Pareto-improving cross-fleet strategy.  This strategy serves as an upper bound of the achievable total profit of all trucks in a batch. The system-maximum strategy is modeled as the following optimization problem:
 \begin{subequations}  \label{coop}
\begin{align}
	&\underset{\epsilon_p: p\in \mathcal C_h(t) }{\text{max}}&&  \sum \limits_{p\in \mathcal C_h(t)}   r_p \epsilon_p   & \\
	&\text{s.t.} &&
	\sum \limits_{p\in \mathcal C_{h,i}(t) }\epsilon_p =1, & \forall i\in \mathcal B_h(t)   \\
	&&& \epsilon_p=\{0,1\}, & \forall p\in \mathcal C_h(t),
	\end{align}
	 \end{subequations}
with parameters and variables defined as those in problem~ \eqref{par}. 

\subsection{Size of optimization problems}\label{subsec_ccc_cc}


The optimization problems in \eqref{par}, \eqref{iso}, \eqref{coop} are integer linear programs. The number of variables and constraints in these problems depends on the number of feasible platoons and trucks in the considered batch.  For example, the number of variables and constraints in the Pareto-improving optimization problem \eqref{par} are  $|\mathcal C_h(t)|$ and $|\mathcal B_h(t)|+~|\mathcal F_h(t)|$, respectively. The number of feasible platoons $|\mathcal C_h(t)|$ depends on the batch size $|\mathcal B_h(t)|$. In particular, $|\mathcal C_h(t)|\leq  2^{|B_h(t)|}$, where equality occurs when any subset of the trucks in $\mathcal B_h(t)$ can form a feasible platoon, making the pairwise feasibility graph a complete graph \footnote{a complete graph is a simple undirected graph in which every pair of distinct vertices is connected by a unique edge.}. This emphasizes the importance of explicitly limiting the number of variables and the number of constraints in the optimization problems. These are limited in the batch selection process, explained in Section \ref{PWFG}, by selecting the batch such that the number of trucks in the batch and the number of feasible platoons are below selected thresholds. In practice, the thresholds are selected to achieve an acceptable computational time.

\section{Simulation study}\label{SSpi}

In this section, we use the Swedish transportation system as the underlying network and evaluate the cross-fleet Pareto-improving coordination strategy by comparisons with single-fleet and system-maximum coordination strategies.  We first explain the simulation setup, then present the evaluation of the numerical results. Apart from the profit obtained by the proposed strategies, we also present the computational time required to compute the coordination solutions and demonstrate the importance of proper batch size selection.

\subsection{Setup}

\begin{figure}
    \centering
    \includegraphics[scale=0.5]{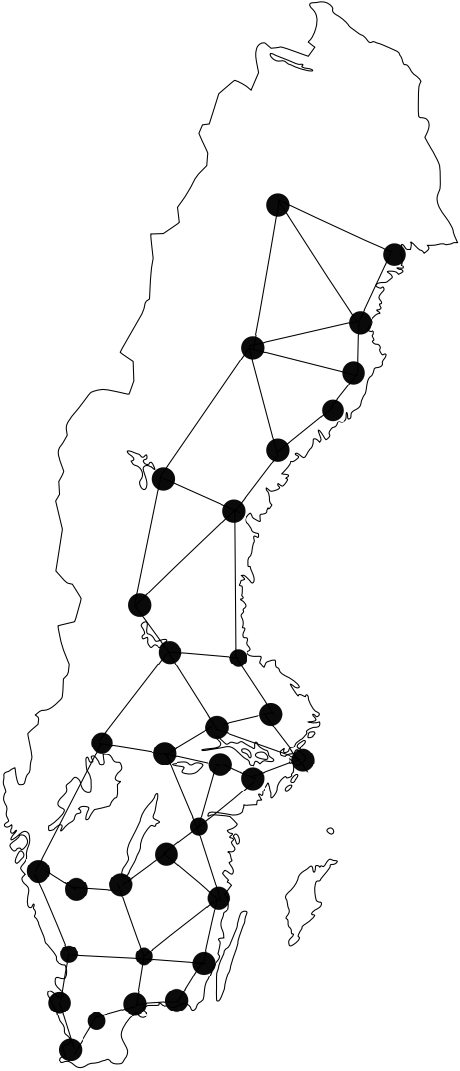}
    \caption{Network of hubs where trucks can wait to form platoons.  }
    \label{RNpi}
\end{figure}

We consider the Swedish transportation network in Figure~\ref{RNpi}, which includes $33$ hubs and $52$ roads connecting the hubs. Each truck's origin and destination are randomly drawn from the set of hubs with an equal probability.  The trucks' paths are their shortest paths between their origins and destinations. Each of the trucks belongs to one of four considered fleets. The percentage of all trucks that belong to the fleets are $40 \%$, $30 \%$, $20 \%$, and $10 \%$, respectively. We will consider the cases when the total number of trucks in the network are $500$ and $2500$, respectively. The trucks have starting times during a period of three hours.

We use the platooning reward function in equation \eqref{PR}, where the platooning reward is shared among the fleets in proportion to the number of trucks from each fleet.  We assume that the platooning reward for each following truck is $ \$ 5.25  /100\text{ km}$. This is accurate if the platooning reward is due to reduced fuel consumption and each following truck saves $10\%$ of fuel.  The waiting cost function in equation \eqref{WC} is used, and we assume that  the cost of waiting is $ \$ 20  $ per hour.  Each truck has a waiting budget of $\bar w_i= 20$ minutes. Coordination is triggered at a hub when an uncoordinated truck will arrive within $\Delta T_{trigger}=5$ minutes. To achieve acceptable computational times, we limit the batch such that the number of trucks is less than  $\bar b=25$ and the number of feasible platoons is less than  $\bar c=6000$.

\subsection{Benefits of cross-fleet platooning}

\begin{figure}
    \centering
    \subfigure[$500$ trucks]
    {
        \includegraphics[scale=0.44]{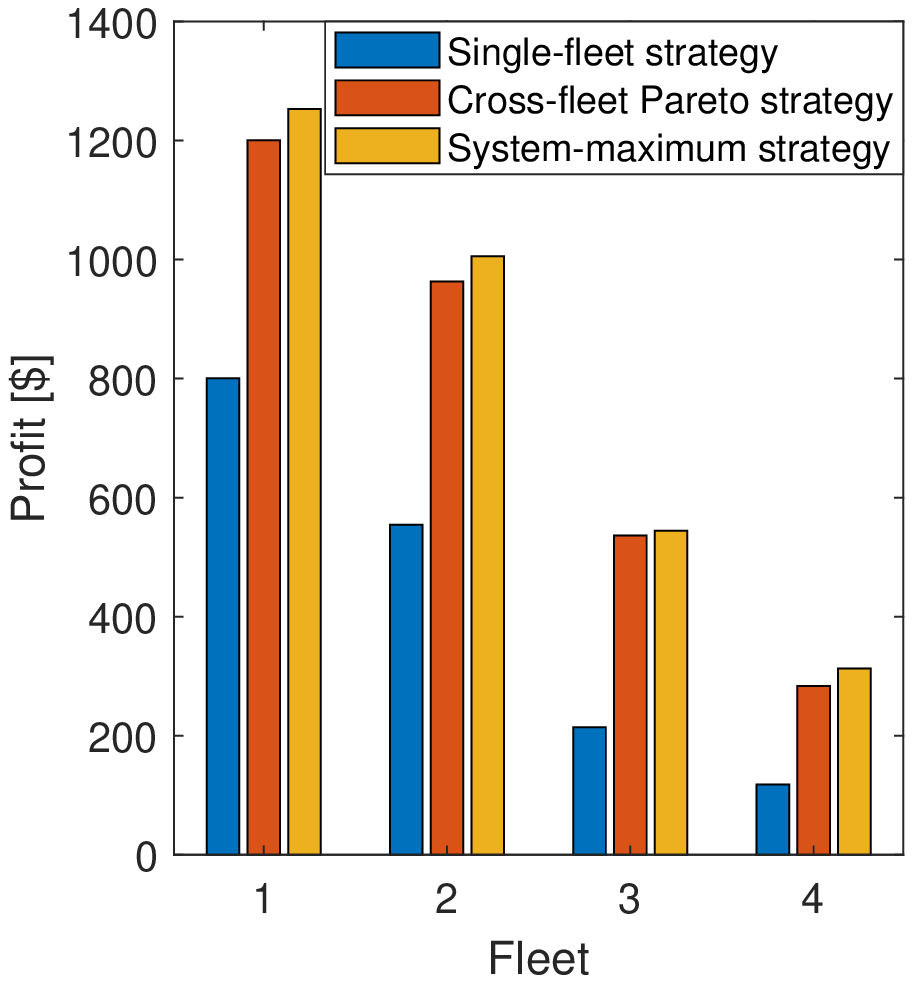}\includegraphics[scale=0.44]{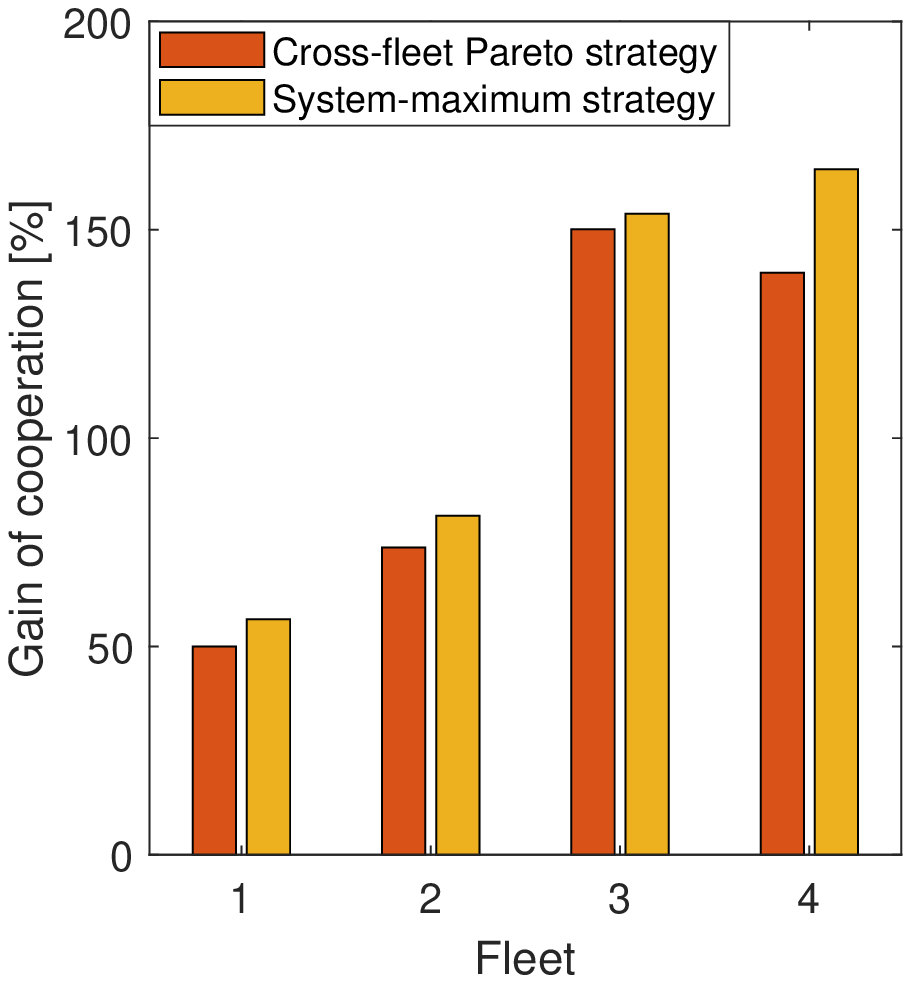}
        \label{Profitagain500a1500_1}
    }
    \subfigure[$2500$ trucks]
    {
        \includegraphics[scale=0.44]{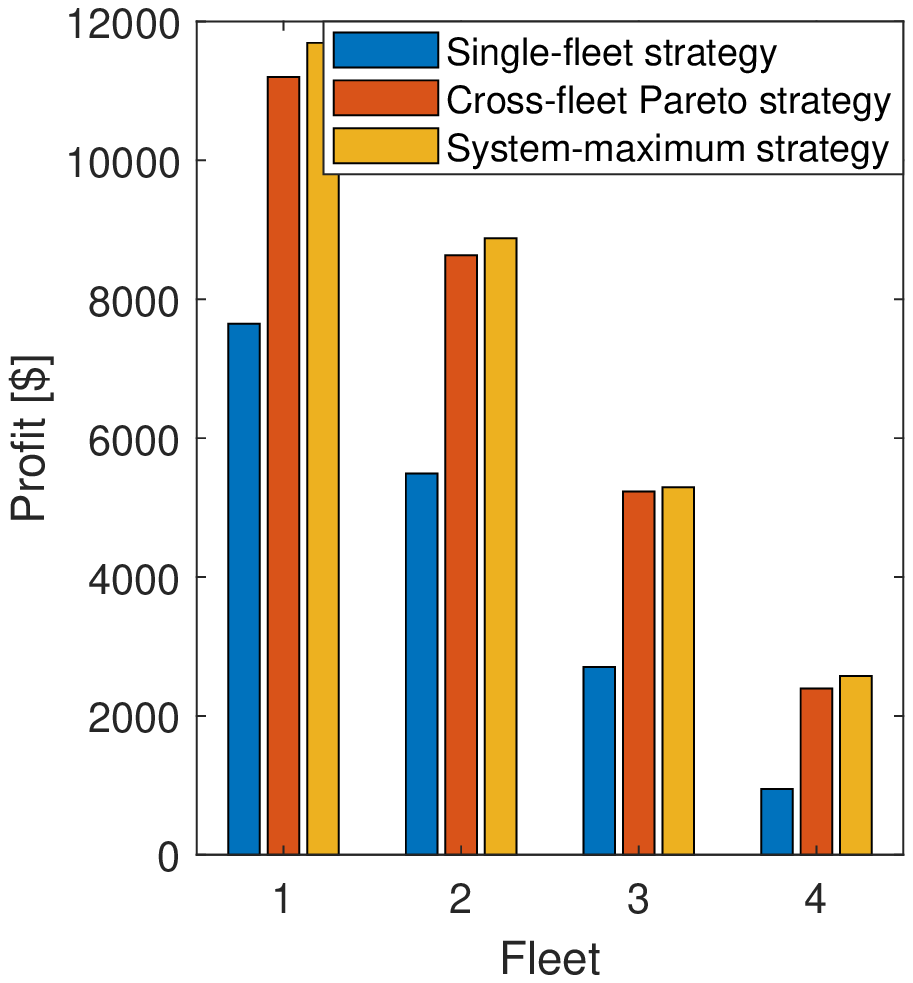}\includegraphics[scale=0.44]{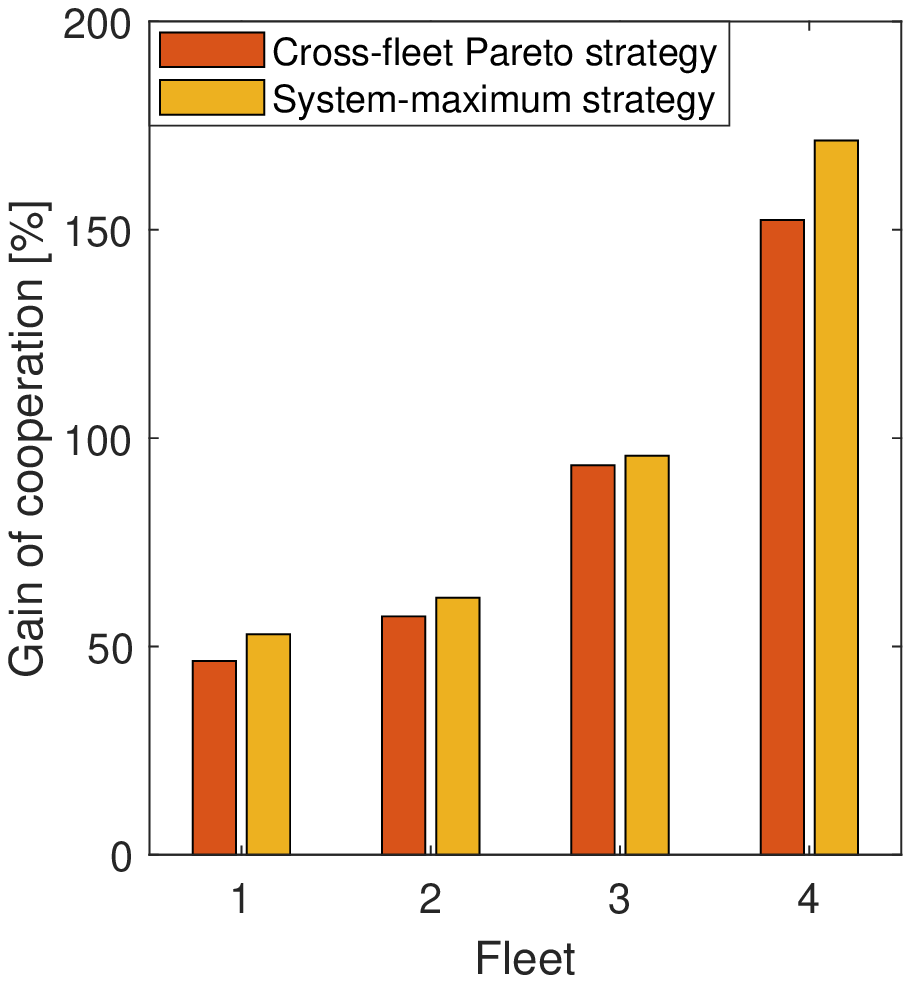}
        \label{Profitagain500a1500_2}
    }
    \caption{Fleets' profits and profit gains due to cross-fleet platooning when the number of trucks is $500$ and $2500$ in sub-figures (a) and (b), respectively. The fleets enumerated $1,2,3$, and $4$ consist of $40 \%$, $30 \%$, $20 \%$, and $10 \%$ of the trucks in the network, respectively.}
    \label{Profitagain500a1500}
\end{figure}

This section compares the fleets' profits and the reduced fuel consumption under cross-fleet Pareto-improving, single-fleet, and system-maximum strategies.  Figure
\ref{Profitagain500a1500} shows the fleets' profits and their profit gains of cross-fleet Pareto-improving and system-maximum strategies from single-fleet strategy. When comparing the profits in Figures
\ref{Profitagain500a1500_1} and \ref{Profitagain500a1500_2}, one can see that the more trucks in the system, the greater platooning profits are. It also can be seen that cross-fleet platooning significantly increases all fleets' profits, especially for the smaller fleets. For example, when the number of trucks is $2500$, the profit increase due to cross-fleet platooning of fleet $1$ and $4$ are $46\%$ and $152\%$, respectively. This is intuitive since larger fleets have more within-fleet platooning opportunities than smaller fleets.

\begin{figure}
    \centering
    \includegraphics[scale=0.6]{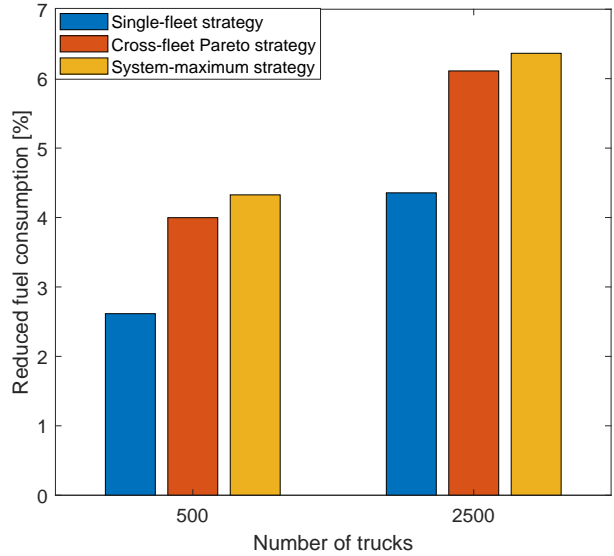}
    \caption{Reduced fuel consumption due to platooning when the number of trucks is $500$ and $2500$, respectively.}
    \label{Redfuel500a1500}
\end{figure}

Figure \ref{Redfuel500a1500} shows the reduced fuel consumption due to platooning, which increases with the number of trucks connected to the coordination system. Cross-fleet platooning can increase the reduction significantly. For example, when the number of trucks in the network is $2500$, single-fleet platooning achieves a reduction of $4 \%$ and cross-fleet Pareto-improving platooning achieves a reduction of nearly $6 \%$. System-maximum platooning achieves a reduction slightly above that of cross-fleet platooning. Note that an upper bound of the reduced fuel consumption due to platooning is $10\%$, which is unattainable as it requires all trucks driving as platoon followers during their entire trips.

\subsection{Computational efficiency}\label{subsec_simu_ce}

\begin{figure}
    \centering
    \subfigure[Computational time as a function of the number of variables]
    {
       \includegraphics[scale=0.6]{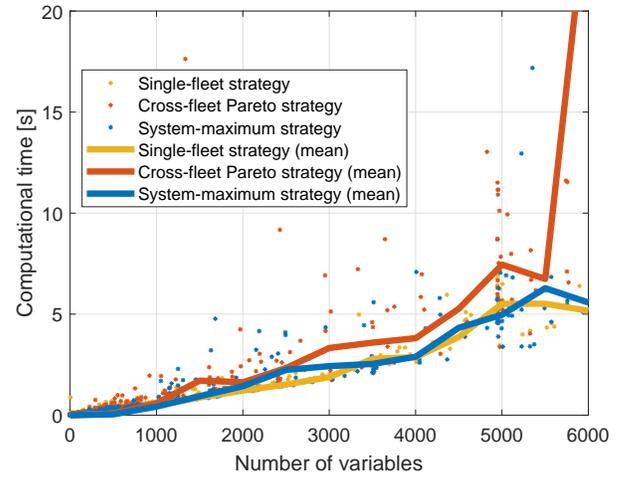}
        \label{Comp1000}
    }
    \\
    \subfigure[Computational time as a function of the batch size]
    {
        \includegraphics[scale=0.6]{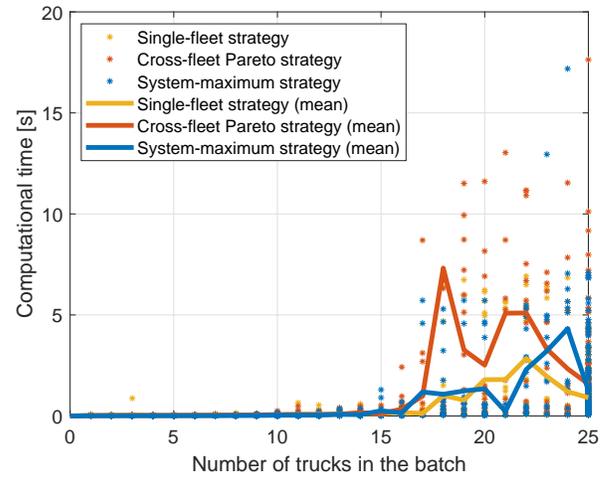}
        \label{Comp10002}
    }
    
    \subfigure[Number of variables as a function of batch size]
    {
        \includegraphics[scale=0.6]{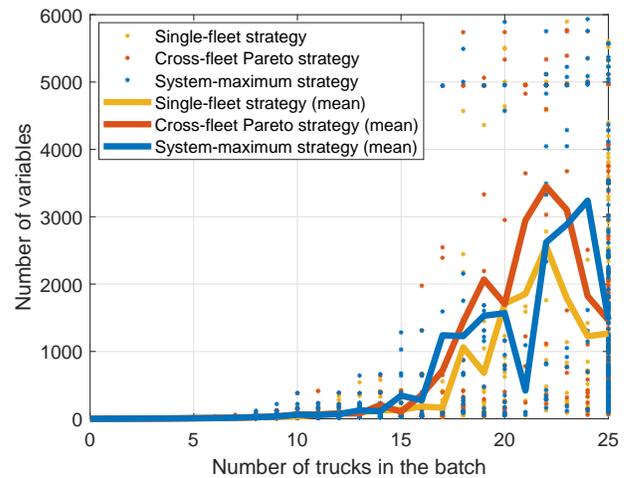}
        \label{Comp10003}
    }
    
    \caption{Computational time as a function of the number of variables (a), computational time as a function of the batch size (b), and the number of variables as a function of batch size (c). One outlier with a computational time of $93$ seconds is outside the range of the plots. The outlier is the reason for the steep rise at $5500$ number of variables in (a) and the peak at $18$ trucks in the batch in (b).}
    \label{Comp10004}
\end{figure}

We demonstrate the computational efficiency of the proposed platoon coordination strategies, which is crucial for real-time platoon coordination.  We evaluate the computational time at coordination instances in the simulation with $2500$ trucks. Figure \ref{Comp10004} shows the computational time with respect to the number of variables and batch size, and the relationship between the number of variables and batch size. Figures \ref{Comp1000} and  \ref{Comp10002} show that the computational time is marginally higher for the cross-fleet Pareto-improving strategy than for the other strategies. The figures indicate that when the number of variables (feasible platoons) is less than $6000$ and the batch size is less than $25$ trucks, the computational time is still less than 20 seconds, except for an outlier of $93$ seconds. Figure \ref{Comp10003} shows that the batch is limited by both the number of trucks and the number of feasible platoons since both entities reach their limits of $\bar b=25$ and $\bar c=6000$ during the simulation.

\section{Conclusions and future work}\label{CFpi}

In this paper, we propose a cross-fleet platoon coordination system that accommodates multiple hubs in the transportation network to provide real-time cross-fleet coordination. The coordination strategy based on Pareto-improvements guarantees that each fleet is better off than performing single-fleet platooning. As the number of variables in the coordination strategy's optimization problem grows with the number of trucks, we explain how the batch of trucks considered for a coordination instance is selected to achieve an acceptable computational time.

 The coordination strategies are compared in a simulation study, which indicates that cross-fleet platooning can increase the profit significantly and that smaller fleets gain more from cross-fleet platooning than larger fleets. This suggests that cross-fleet platooning may be essential for smaller fleets to invest in platooning technology.  Compared with the system-maximum strategy, the cross-fleet Pareto-improving strategy also achieves most of the total profits, making the envisioned cross-fleet platoon coordination viable. The simulation study also demonstrates that cross-fleet platooning can significantly reduce the environmental impact of trucks.

There are several possible directions for our future research. In this paper, many parameters for the dynamic coordination process, including the hub location, the trigger time, and the batch size, are chosen empirically. One avenue of future research is studying their relationships theoretically and optimizing them to improve system performance while guaranteeing computational efficiency. Adding stochasticity on trucks' travel times and the flexibility to change paths into the current framework would be another interesting direction to explore. It should  also be noted this paper focuses on the scenario where information sharing among fleets is difficult and proposes distributed decision-making process. However, in the future, we will compare the distributed platoon coordination system with a global platoon coordination system that has information about all trucks in the network.

\bibliographystyle{IEEEtran}

\bibliography{RefDatabase}

\vspace{12pt}

\end{document}